\documentclass[twocolumn]{aastex631}

\usepackage{amsmath}
\usepackage[caption=false]{subfig}
\usepackage{bm}

\newcommand\TV{\text{TV}}
\newcommand\mcL{\mathcal{L}}
\DeclareMathOperator*{\argmin}{arg\,min}

\shorttitle{A Poisson Process AutoDecoder for X-ray Sources}
\shortauthors{Song et al.}

\graphicspath{{./}{figures/}}
\begin{document}

\title{A Poisson Process AutoDecoder for X-ray Sources}

\author{Yanke Song}
\affiliation{Department of Statistics, Harvard University}

\author{V.~Ashley Villar}
\affiliation{Center for Astrophysics $|$ Harvard \& Smithsonian, Cambridge, MA 02138, USA}
\affiliation{The NSF AI Institute for Artificial Intelligence and Fundamental Interactions}

\author{Rafael Martínez-Galarza}
\affiliation{AstroAI}
\affiliation{Center for Astrophysics $|$ Harvard \& Smithsonian, Cambridge, MA 02138, USA}

\author{Steven Dillmann}
\affiliation{AstroAI}
\affiliation{Institute for Computational and Mathematical Engineering, Stanford University}

\begin{abstract}
%X-ray observing facilities such as the Chandra X-ray Observatory and the XMM-Newton telescope have detected hundreds of thousands of sources associated with high energy phenomena. More recent and ongoing surveys, such as the eROSITA all-sky survey, have increased the total number of serendipitous sources by about an order of magnitude. 
%The arrival of photons as a function of time follows a Poisson process and can vary by orders-of-magnitude in length, presenting obstacles for downstream tasks such as source classification, physical property derivation and anomaly detection. Previous work has either failed to directly capture the Poisson nature of the data, or requires complex pipelines that make strong assumptions regarding the spectral response of the instrument, the shape of the PSF function, and the statistical properties of the underlying Poisson rate functions (light curves). 
%Here, we present a Poisson Process AutoDecoder (PPAD) to learn the time-variable, multi-band X-ray light curves. PPAD is a neural field decoder that maps fixed-length latent features to continuous light curves in an unsupervised manner. It reconstructs the rate function and yields a representation at the same time. As a proof-of-concept, we use PPAD to learn sources in the Chandra Source Catalog and demonstrate its effectiveness in light curve reconstruction, regression, classification and anomaly detection.

X-ray observing facilities, such as the Chandra X-ray Observatory and the eROSITA, have detected over a million astronomical sources associated with high-energy phenomena. The arrival of photons as a function of time follows a Poisson process and can vary by orders-of-magnitude, presenting obstacles for common tasks such as source classification, physical property derivation, and anomaly detection. Previous work has either failed to directly capture the Poisson nature of the data or only focuses on Poisson rate function reconstruction. In this work, we present Poisson Process AutoDecoder (PPAD). PPAD is a neural field decoder that maps fixed-length latent features to continuous Poisson rate functions across energy band and time via unsupervised learning. PPAD reconstructs the rate function and yields a representation at the same time. We demonstrate the efficacy of PPAD via reconstruction, regression, classification and anomaly detection experiments using the Chandra Source Catalog.

\end{abstract}

%% Keywords should appear after the \end{abstract} command. 
%% The AAS Journals now uses Unified Astronomy Thesaurus concepts:
%% https://astrothesaurus.org
%% You will be asked to selected these concepts during the submission process
%% but this old "keyword" functionality is maintained in case authors want
%% to include these concepts in their preprints.
\keywords{}

\section{Introduction}\label{subsec:intro_unsupervised}

X-ray astronomy, like many subfields of observational astrophysics, has entered a new era of ``Big Data". Massive volumes of X-ray data are being produced at unprecedented rates thanks to ongoing X-ray surveys and missions, such as the Chandra X-ray Observatory \citep{evans24csc}, the XMM-Newton \citep{webb20xmmcat} telescope, and the eROSITA survey \citep{merloni23erass}, which together contain approximately 2 million individual X-ray sources in the sky (and several million individual detections). Automatic data processing, analysis and learning has become increasingly more demanded as it enables various downstream applications at massive scale, such as classification of unlabeled sources, rapid identification of high-energy transients and spectral anomalies, as well as scientific evaluation of serendipitous detections \citep{2024MNRAS.tmp.2687D}. However, X-ray sources vary by orders-of-magnitude in terms of X-ray photons detected, as well as in the distribution of photon energies and relevant timescales. Many sources are well-within the Poisson limit--with telescopes receiving just a few photons per exposure per source--thereby posing additional challenges. Machine learning methods have gained popularity recent years as a powerful type of approaches for automated X-ray analysis. Although supervised learning methods have found success in classification tasks \citep{lo2014automatic,farrell2015autoclassification, yang2022classifying}, they require real labels for training, which many X-ray sources lack. Here, we instead focus on unsupervised learning methods due to its label-free property and flexibility for downstream analysis. To give a complete picture, we also include unsupervised methods for sources with available multi-wavelegth data, as many ideas are potentially transferrable for X-ray sources.

A general unsupervised learning framework consists of (1) collecting a set of features, (2) performing optional dimensionality reduction, and finally (3) conducting ``downstream tasks" such as clustering, anomaly detection and classification on the low-dimensional feature embeddings. Previous works can be broadly categorized by how they handle feature extraction. One line of work utilizes descriptive variables---often high level summary statistics---that are extracted from analysts from individual data observations.Examples of these in X-ray astronomical analysis are spectral hardness ratios and variability summaries. These features are then passed to different unsupervised learning algorithms for dimension reduction and/or clustering, such as self-organizing maps \citep{kovavcevic2022exploring}, Gaussian Mixture Models (GMMs; \citealt{perez2024unsupervised}), Density-Based Spatial Clustering of Applications with Noise (DBSCAN; \citealt{giles2019systematic}), Hierarchical DBSCAN + t-distributed Stochastic Neighbor Embedding (t-SNE; \citealt{webb2020unsupervised}), GMM + t-SNE \citep{bhardwaj2023grb}, among others. However, manual feature engineering requires specialized knowledge and may lead to biased feature selection.

Another line of work instead uses the less-preprocessed form of data and attempts automated (i.e., data-driven) feature extraction and low-dimensional embedding. Although traditional machine learning methods have been used in such settings \citep{armstrong2015k2, mackenzie2016clustering, valenzuela2018unsupervised}, neural networks often find success in this more challenging task of extracting patterns without manual features. For example, \cite{naul2018recurrent} and \cite{chan2022searching} use a recurrent neural network (RNN) and convolutional neural network (CNN), respectively, to extract features of folded light curves of variable sources, whereas \cite{orwat2022light} and \cite{ricketts2023mapping} use Long Short-Term Memory (LSTM) to extract features of segments of a large light curve on GRS1915+105. 
Moreover, due to its superior representation learning ability, neural networks trained on supervised tasks often learn informative embeddings in their hidden layers. In this regard, end-to-end architectures for supervised tasks also serve unsupervised learning purposes, and previous works have explored different neural network architectures on this line, such as RNN \citep{becker2020scalable, villar2020superraenn}, bi-directional RNN \citep{charnock2017deep}, CNN \citep{shallue2018identifying} and Cyclic-Permutation Invariant Network \cite{zhang2021classification}, among others. However, all methods mentioned above focuses on optical light curves, for which the abundance of photons are well within the large-number Gaussian limit and the stochastic arrivals with Poisson nature can be ignored. 

To transfer these ideas to X-ray data, one needs to reconstruct the light curves for X-ray sources. There exists a robust and Bayesian approach for X-ray light curve reconstruction, known as the Gregory-Loredo algorithm \cite{gregory1992new}. Specifically, it proposes a uniform prior on light curve hypotheses (usually stepwise ones), combines the prior with Poisson Process likelihoods, and obtains the posterior probabilities for different light curves. It then superimposes the hypotheses weighted by posterior probabilities to obtain the reconstructed light curve. However, the GL algorithm only considers stepwise hypotheses (often with less than $20$ steps) due to its intense computational complexity, thereby limiting the resolution of the reconstruction. More importantly, the reconstructed light curves from the GL algorithm need further analysis for feature extraction. Instead, an ideal unsupervised learning framework would be capable of extracting features in an end-to-end fashion, directly from the event files themselves (i.e., the arrival times and energies of these events). \cite{2024MNRAS.tmp.2687D} is one of the first works along this line, proposing to use a sparse autoencoder on energy-time binned histograms of event files for automatic feature extraction, for which resulting features can be directly used by t-SNE and DBSCAN for further dimension reduction and clustering. Binning the event files, however, ignores the intrinsic stochastic nature of photon arrivals, thereby potentially creating artifacts which are especially severe for low-count sources.

In this work, we propose the Poisson Process AutoDecoder (PPAD), a pipeline that embeds \textit{raw} event files to latent representations in an unsupervised manner. 
PPAD addresses the aforementioned challenges by making three significant contributions. First, it employs a neural field for light curve reconstruction, offering continuous resolution and bypassing the binning in previous approaches. Second, it uses a Poisson likelihood-based approach that respects the intrinsic stochasticity of X-ray sources. Third, via an autodecoder, it learns fixed-length latent representations of variable-length event files, offering great flexibility for downstream tasks.

Our light curve reconstruction method employs a one-dimensional neural field, which has gained tremendous popularity in the machine learning community, especially in 2D/3D computer vision \citep{park2019deepsdf, mildenhall2021nerf}. A neural field implicitly represents a signal via a neural network, and enjoys distinct advantages such as continuity and memory efficiency. In the context of light curve representation, instead of using a fixed-length vector to explicitly represent a light curve via its intensity at a series of time-steps, we choose to represent a light curve using a neural network, which represents an implicit function that maps any time value to the light curve intensity, thereby making it resolution-free. The output light curve is then compared to the raw event file data and a Poisson likelihood based loss function used to optimize the neural field representation. We also employ techniques such as positional encoding and total variation penalty to improve the reconstruction quality.

To enable joint learning from a collection of event files, we utilize an autodecoder approach. Specifically, a shared neural network is used to reconstruct all light curves, except that one unique fixed length latent vector is added as an extra condition to each event file. These latent vectors are optimized together with the neural network. When training is completed, not only do we get reconstructed light curves for respective sources, but we also obtain these latent vectors as low-dimensional representations of these light curves that are useful for downstream tasks.

The rest of this paper is structured as follows: Section \ref{sec:data_processing} describes our data processing pipeline, which retains raw event files of the Chandra Source Catalog. Section \ref{sec:method} describes techniques and motivations of our main method in detail. Section \ref{sec:experiments} presents experimental results that showcase the functionality of our method in light curve reconstruction, source classification and anomaly detection. Finally, Section \ref{sec:discussion} summarizes our results, discusses limitations, and articulates directions of future research.

\section{Data and Preprocessing}\label{sec:data_processing}
We utilize data from the Chandra Source Catalog, CSC \citep{evans24csc} to train and test our PPAD algorithm. The data is in the form of \emph{event files}, which are data structures containing individual X-ray photon recordings associated with a single astrophysical X-ray source in the sky. Event files can be understood as multivariate time series of the photon's energies, their coordinates on the detector, and other relevant quantities. The energies of the recorded photons cover a range between approximately 0.5 keV up to about 8 keV. X-ray properties of astrophysical sources, such as their spectral hardness and variability probability, are computed as summary statistics from these event files and compiled in the CSC, together with many other quantities. Of relevance for this paper are the following X-ray properties:

\begin{itemize}
\item \emph{Hardness ratios}: A quantification of the distribution of photon energies between three energy bands: soft (0.5~keV-1.2~keV), medium (1.2~keV-2~keV), and hard (2~keV-7~keV). They are broadly defined as the difference in X-ray flux between two bands, divided by their sum. This information is relevant for assessing the physical mechanism (e.g., thermal versus non-thermal) producing the X-ray emission. In the CSC, hardness ratios are represented by the properties \verb|hard_hs|, \verb|hard_ms|, and \verb|hard_ms|.

\item \emph{Variability probability}: The probability that the photon arrival times, understood as a Poisson process, are consistent with a change in the Poisson rate as a function of time. It is computed using the Gregory-Loredo algorithm \citep{gregory1992new}. This quantity is of relevance to severe changes in the physical conditions, such as explosive events or variations in the accretion flows towards compact objects. In the CSC, the probability that an X-ray detection is variable in the integrated (broad) energy band is represented by the property \verb|var_prob_b|.

\item \emph{Variability index}: A measure of the confidence at which variability (the previous quantity) is determined. It is computed from the odds that the photon arrival times can result in the observed binned values in the absence of true variability. In the CSC, the variability index for the integrated (broad) energy band is represented by the property \verb|var_index_b|.
\end{itemize}

We use the event files dataset from \cite{2024MNRAS.tmp.2687D}, which contains $\sim$100,000 event files from the CSC. We employ the following pre-processing:
\begin{itemize}
    \item Energies are binned in soft ($E \in [0.5,1.2]$ keV), medium ($E \in [1.2,2]$ keV), and high ($E \in [2,7]$ keV) light curve bins in order to minimize computational cost of the loss function. However, we note that this step is not necessary and (as we will show in Section \ref{subsec:PPAD}), our method in principle supports finer binning.
    \item Event files are truncated to have the same lifetime of $8$ hours. Event files shorter than $8$ hours are omitted and those longer than $16$ hours are truncated into multiple separate event files.
    \item All event files are normalized so that first arrival happens at time $0$.
\end{itemize}

After pre-processing, our dataset contains 109,656 event files, each $8$ hours long.

\section{Architecture and Training}\label{sec:method}
\subsection{Modeling Photon Arrivals as Poisson Processes}\label{subsec:likelihood}
Here, we describe the statistical framework in which we consider each source in our training set. For simplicity, in the following description we ignore the X-ray photon energies, but as we will demonstrate later, the following principles hold equally for energy-time series in the event file.
It is common practice \citep{cash1979parameter} to model stochastic photon arrivals in an event file as a Poisson process. In order to capture the underlying physical change of X-ray sources (non-constant light curve), we will use the more general inhomogeneous Poisson processes. It is well known that, for an inhomogeneous Poisson process with rate $r$ (effectively the light curve intensity), the likelihood of a list of photon arrivals $\{t_i\}_{i=1}^n$ during an observation interval $[0,T)$ is (see, e.g. \citealt{rasmussen2018lecture}):

\begin{equation}\label{eqn:likelihood}
\text{likelihood}(t_1,...,t_n;r) = \left(\prod_{i=1}^n r(t_i)\right) \exp\left(-\int_0^T r(t) dt\right).
\end{equation}

Here the integral is approximated via $N$ uniformly discretized points in $[0,T)$: 

\begin{equation}
    \tau_i = \frac{i-1}{N-1} T.
\end{equation}

Given a list of events $\{t_i\}_{i=1}^n$ on $[0,T)$, we would like to find the light curve $r$ that maximizes the likelihood--or equivalently minimizes the negative log-likelihood--of this event file. However, this is an ill-posed problem. A straightforward check reveals that a light curve with large values at arrival times $\{t_i\}_{i=1}^n$ and zero values elsewhere yields unbounded log-likelihood. Therefore, we need additional constraints to regularize the problem. 

We want the regularization term to have the following desired properties: 1) It penalizes the change rate of the light curve instead of the raw value itself, since different sources might naturally have variations in base rates; and 2) Instead of imposing smoothness, it encourages sparsity and piecewise constancy, since a source might undergo abrupt change of rates during transient behaviors but retains a relatively constant rate otherwise. Our regularization term does \textit{not} require analytical derivatives, since we will fit these rate functions via neural networks (see Section \ref{sec:neural_field}). Based on these criteria, we choose the discretized total variation--hereafter simply referred as the total variation (\TV)--as the additional penalty term. Specifically, for the set of discretization points $0  = \tau_1 \leq ... \leq \tau_N = T$, the total variation of the rate function $r(t)$ on these points is defined as:
\begin{equation}\label{eqn:def:TV}
\TV(r;\tau_1,...,\tau_N) = \frac{1}{N-1}\sum_{i=1}^{N-1}|r(\tau_i)-r(\tau_{i+1})|.
\end{equation}

Applying the total variation penalty only on the set of discretization points, however, does not provide sufficient regularization on the rate function at arrival times $\{t_i\}_{i=1}^n$. Therefore, we apply an additional total variation loss on the arrival times to make sure that the penalty is also adequately sampled at high-count regions.

Summing up the negative log-likelihood and the total variation penalties, the loss for a given light curve $r$ is given by:

\begin{equation}\label{eqn:loss_single}
\begin{aligned}
\mcL(r) :=& l_{\text{likelihood}} + l_{\text{TV}}\\
=& -\sum_{i=1}^n \log r(t_i) + \int_{0}^T r(t) dt \\
&+\lambda_{\text{TV}} \bigg[\frac{1}{N-1}\sum_{i=1}^{N-1}|r(\tau_i)-r(\tau_{i+1})| \\
&+ \frac{1}{n-1}\sum_{i=1}^{n-1}|r(t_i)-r(t_{i+1})|\bigg], 
\end{aligned}
\end{equation}
where we have dropped the dependence on $\{t_i\}_{i=1}^n$ and $\{\tau_i\}_{i=1}^N$ for conciseness. Here, $\lambda_{\text{TV}}$ is a hyperparameter that adjusts the total variation penalty level.

\subsection{Neural Representation of Light Curves}\label{sec:neural_field}

In order to find the light curve that minimizes the loss $\mcL(r)$, we choose to parameterize $r$ via a neural network--hereafter referred to as the neural representation. Neural networks are a key component in modern deep learning practice and have proved powerful in approximating complex signals \citep{lu2017expressive}. We can then use standard gradient descent algorithms (e.g., Adam \citealt{kingma2014adam}) to minimize the loss defined in Eqn. \ref{eqn:likelihood} by tuning $\phi$. Upon convergence, $r_\phi$ yields the reconstructed light curve of the given event file.

The canonical approach of neural representation is to let the neural network output a discretization of the signal (e.g. a convolutional neural network outputs a fixed-resolution image), partly because most signals are already discrete when collected. In our setting, such a neural network would output a $d_{\text{out}}$ dimensional vector, representing the value of $r$ at $d_{\text{out}}$ discretized points. There are two drawbacks with this canonical approach: 1) To capture high-frequency signals such as transients, we need dense discretization. However, a large $d_{\text{out}}$ means a larger network and more computational overhead. 2) The architecture of such a neural network is tied to its $d_{\text{out}}$, limiting its resolution and flexibility. Instead, we choose to use a neural network to directly model the function $r$ itself. In other words, the neural network (with weights $\phi$) would take time $t$ as an input and output $r_{\phi}(t)$ such that $r_{\phi}(t) \approx r(t)$. This is known as the neural field representation and is now common practice in recent machine learning literature to represent spatial signals (e.g. \citealt{mildenhall2021nerf}). 
The advantages of this representation lies in its ability to continuously represent a signal, therefore allowing efficient computation and flexible adaptation.
%One would then proceed to calculate the loss $\mathcal{L}(r_\phi)$ defined in \eqref{eqn:loss_single}, back-propagate the loss, and apply gradient descent to optimize for the network weights $\phi$. Given that the training converges, $r_\phi$ would yield the reconstructed light curve of the given event file.

\subsubsection{Positional Encoding}
Although neural networks are known to be universal function approximators \citep{lu2017expressive}, there exist tricks that enhance training efficiency in practice. Specific to our setting, we would like the neural networks to  learn patterns of different frequency, from constant rates to low-frequency variations and high-frequency transients. To this end, we apply Positional Encoding (PE) to the input $t$ before passing it to the neural network. PE is a set of deterministic sinusoidal encodings that first appeared in transformer-based architectures \citep{vaswani2017attention}, but later proved crucial for continuous neural representations \citep{mildenhall2021nerf}. Formally, the encoding function we use is
\begin{equation}\label{eqn:positional_encoding}
\gamma(t) = [\bar{t}, \sin(2^0 \pi \bar{t}), \cos(2^0 \pi \bar{t}),...,\sin(2^{L-1} \pi \bar{t}),\cos(2^{L-1} \pi \bar{t})]
\end{equation}
for $\bar{t}=t/T$. $\gamma(t)$ maps $t$ to a $(2L+1)$-dimensional vector $\gamma(t)$ with features of different frequencies, which is then fed into the neural network to produce the output $r_{
\phi}(\gamma(t))$. Besides creating features of different frequency, the PE also standardizes $t$ into values in $[0,1]$, both of which greatly help increase the expressive power of neural networks.

\subsection{Learning the Latent Space of Features}
Up to now, we have managed to reconstruct the light curve of a single event file using a Poisson likelihood-based loss function and a neural representation. However, effective unsupervised learning necessitates a common feature space where we can compare different sources/event files. Therefore, instead of training a specific neural network for each event file, we want a model that is capable of representing a wide variety of rate functions, discover their similarities/differences, and yield embeddings which are useful for downstream tasks. To this end, we propose to represent each event file via a latent vector $\bm{z}$, and learn these latent values (``latents") together with the aforementioned rate functions using a common neural network.

\subsubsection{Encoder-less Learning}
When it comes to learning neural latent variable representations, autoencoders (and their variants) are one of the most commonly employed architectures. Canonical autoencoders learn to reconstruct the data via an encoder and a decoder that are connected by a lower dimensional bottleneck layer. This forces the neural network to learn lower dimensional abstract representations of the data that are useful for downstream tasks.
Despite their popularity and effectiveness, autoencoders are not appropriate for event files learning in our context. Compared to the canonical autoencoder training where one aims to reconstruct the input data, we aim to reconstruct the light curve from raw event files, resulting in a mismatch between inputs and outputs. Furthermore, compared to time series data and text data where RNN often finds success, Poisson arrival times in event files have much lower signal-to-noise ratio and much higher variance in information throughput. 
Therefore, we instead adopt an autodecoder architecture, which has also become popular in the machine learning literature where encoders are hard to train \citep{sitzmann2019scene, park2019deepsdf}.

\begin{figure}[h]
\vspace{0.1in}
\centering
\subfloat[Autoencoder]{%
    \includegraphics[width=0.23\textwidth]{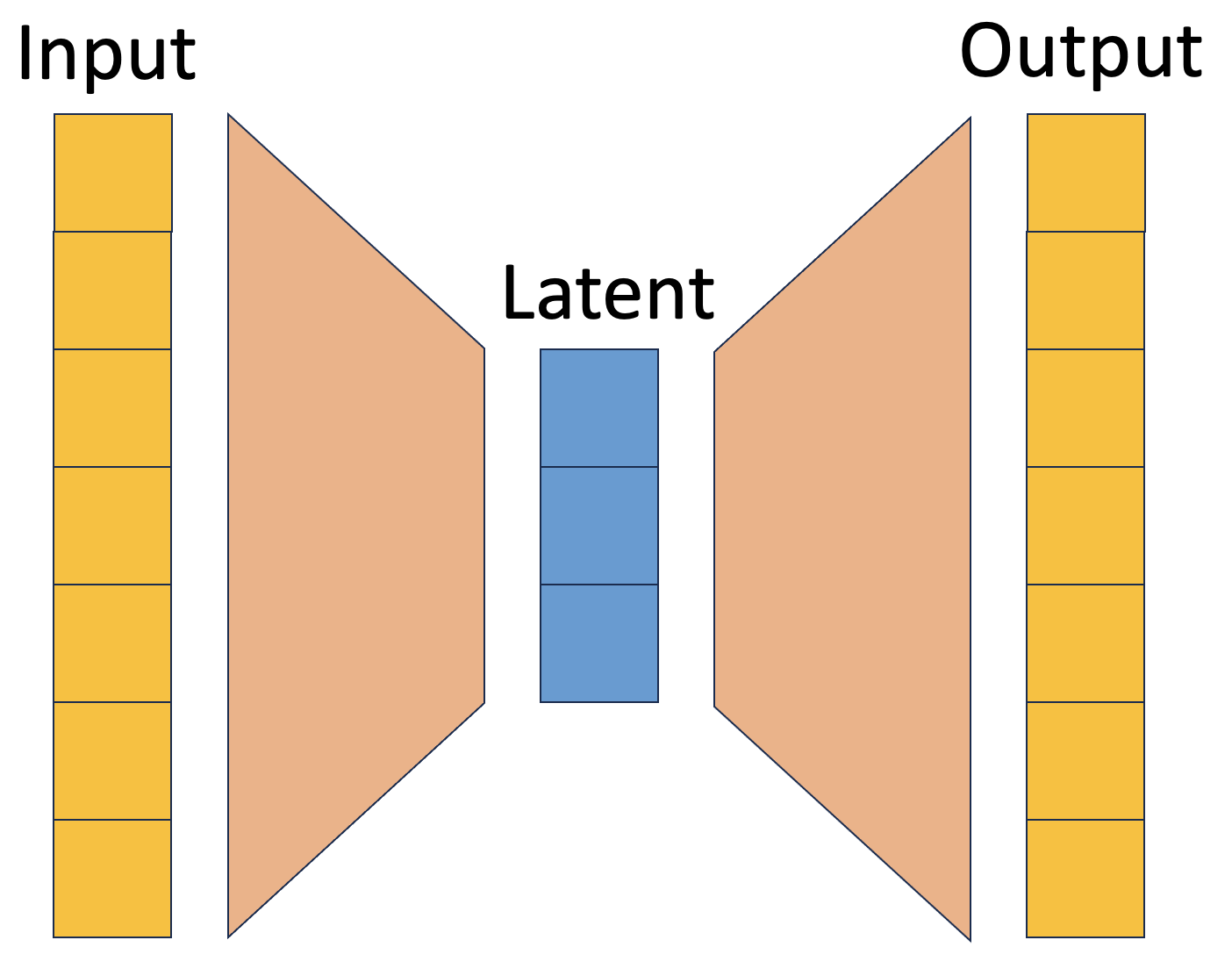}
}
\hfill
\subfloat[Autodecoder]{%
    \includegraphics[width=0.23\textwidth]{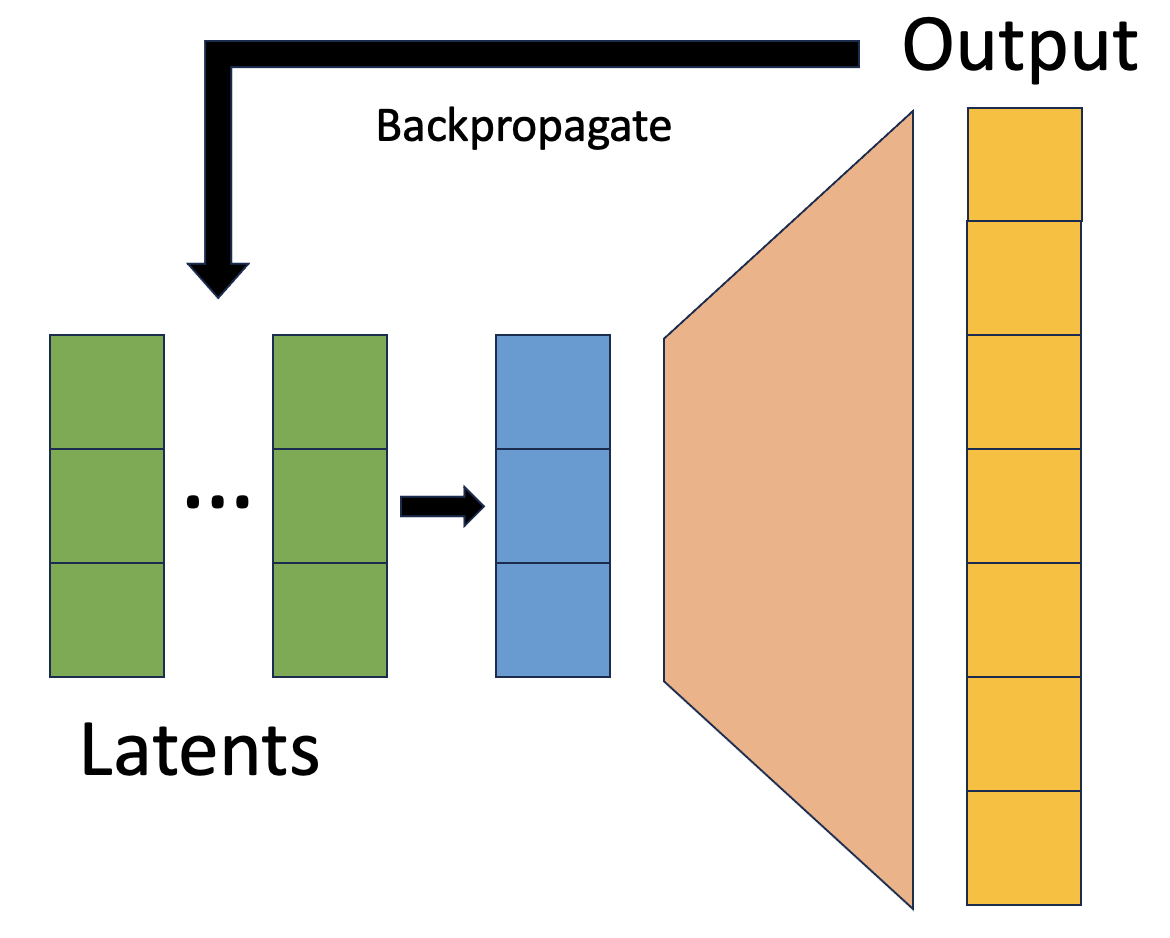}
}
\caption{Compared to an autoencoder where the latent vectors are produced by the encoder, an autodecoder directly accepts latent vectors as inputs. A randomly initialized latent vector is assigned to each data point (event file) in the beginning of training, and latent vectors are optimized together with the decoder weights through gradient descent. At inference time on a new data point, decoder weights are frozen, and a new latent vector is optimized via gradient descent.}
\end{figure}

In an autodecoder, latent variables are directly prepared instead of obtained from an encoder. Specifically, to represent a rate function via a neural network, we randomly initialize a latent variable $\bm{z}$, which is fed together with the PE $\gamma(t)$ through the neural network $r_{\phi}$ to produce the reconstructed light curve. The latent $\bm{z}$ can be viewed as an extra condition that indicates the identity of the neural light curve. 
For a set of event files $\{t_{ji}\},1\leq j\leq m,1\leq i \leq n_j$ coming from $m$ sources, we reconstruct $m$ light curves $r^{(j)}(t) \approx r_\phi(t;\bm{z}^{(j)})$ with the same neural network $\phi$ and different latent variables $\bm{z}^{(j)},1\leq j \leq m$. The set of latents are optimized together with the neural network weights during training. Once trained, the latents $\{\bm{z}^{(j)}\}_{j=1}^m$ become learned representations of the light curves reconstructed from event files, which can be used for downstream tasks. During training, the autodecoder learns information about the full distribution of reconstructed light curves, allowing for generalization to unseen data. At test time, given a previously unseen event file, the weights $\phi$ are frozen and a latent $\bm{z}$ is optimized for the file.

%Although one only obtains the latent variables of event files seen at training time, the neural network has stored information of the whole distribution of reconstructed light curves, which allows generalization to unseen data. At test time, given an unseen event file, one can freeze the weights $\phi$ and optimize a new latent $\bm{z}$, which serves as the representation of the new event file that is compatible with learned representations for downstream tasks.

To encourage concentration of latents, we impose a penalty on the norm of the latents $\|\bm{z}^{(j)}\|_2^2$. This ensures a compact manifold in latent space and helps with the convergence of results. Equivalently, this can also be viewed as imposing an zero-mean isotropic Gaussian prior distribution on the latent variables. 

\subsection{Putting it together: Poisson Process AutoDecoder}\label{subsec:PPAD}
We now present our final full pipeline: Poisson Process AutoDecoder (PPAD).
Combining previous displays, the loss function of PPAD contains three parts: likelihood, total variation penalty, as well as a latent norm penalty. Moreover, recall that we have ignored energy marking. Fortunately, the formulation allows direct extension to discrete energy binning, since we can effectively reconstruct a different rate function for each energy bin. Summarizing all these components, our final loss function is as follows:

\begin{equation}\label{eqn:loss_final}
\begin{aligned}
&\mathcal{L}_{\text{total}}(\phi;\{\bm{z}_j\}
_{j=1}^M) = \sum_{j=1}^M \left(\sum_{k=1}^K \left(\mathcal{L}_{\text{likelihood}}^{(j,k)} + \mathcal{L}_{\text{TV}}^{(j,k)}   \right) + \mathcal{L}_{\text{latent}}^{(j)}\right),\\
&\mathcal{L}_{\text{likelihood}}^{(j,k)} = -\sum_{i=1}^{n_{j,k}} \log r^{(k)}_{\phi}(\gamma(t_{i,k});\bm{z}^{(j)}) + \int_0^T r_{\phi}^{(k)}(\gamma(t);\bm{z}^{(j)}) dt,\\
&\mathcal{L}_{\text{TV}}^{(j,k)} = \lambda_{\text{TV}}\bigg[\frac{1}{N-1}\sum_{i=1}^{N-1}|r_\phi^{(k)}(\gamma(\tau_i);\bm{z}^{(j)})-r_\phi^{(k)}(\gamma(\tau_{i+1});\bm{z}^{(j)})| \\
&+ \frac{1}{n-1}\sum_{i=1}^{n-1}|r_\phi^{(k)}(\gamma(t_i);\bm{z}^{(j)}) - r_\phi^{(k)}(\gamma(t_i);\bm{z}^{(j)})|\bigg],\\
&\mathcal{L}_{\text{latent}}^{(j,k)} = \lambda_{\text{latent}} \|\bm{z}^{(j)}\|_2^2,
\end{aligned}
\end{equation}
where $j=1,...,M$ refers to event files; $k=1,...,K$ refers to energy bins; $t_i,i=1,...,n_j$ refers to photon arrivals; $\tau_i,i=1,...,N$ refers to evenly discretized points; and $\gamma$ is the positional encoding defined in Eqn.~\eqref{eqn:positional_encoding}.

During training, $\phi$ and $\{\bm{z}_j\}
_{j=1}^M$ are optimized together:

\begin{equation}\label{eqn:training_optimization}
\hat\phi,\{\hat{\bm{z}}^{(j)}\}_{j=1}^M := \argmin_{\phi;\{\bm{z}_j\}_{j=1}^M}\mathcal{L}_{\text{total}}(\phi;\{\bm{z}^{(j)}\}
_{j=1}^M).
\end{equation}

At test/inference time for a new event file, $\phi$ is frozen and only a new latent $\bm{z}$ is optimized:

\begin{equation}\label{eqn:testing_optimization}
\hat{\bm{z}} := \argmin_{\bm{z}}\mathcal{L}_{\text{total}}(\hat\phi;\bm{z}).
\end{equation}

The neural network $\phi$ is a ResNet \citep{he2016deep} which takes a $(d_{\text{latent}}+2L+1)$-dimensional input (concatenation of the latent vector and the positional time encoding) and outputs a $K$-dimensional vector representing the rate function at $K$ energy bins. Details on the architecture, the hyperparameters $\lambda_{\text{latent}},\lambda_{\text{TV}}$ and other training details can be found in Appendix \ref{sec:implementation}. 

\begin{figure}[h]
\centering
\includegraphics[width=0.475\textwidth]{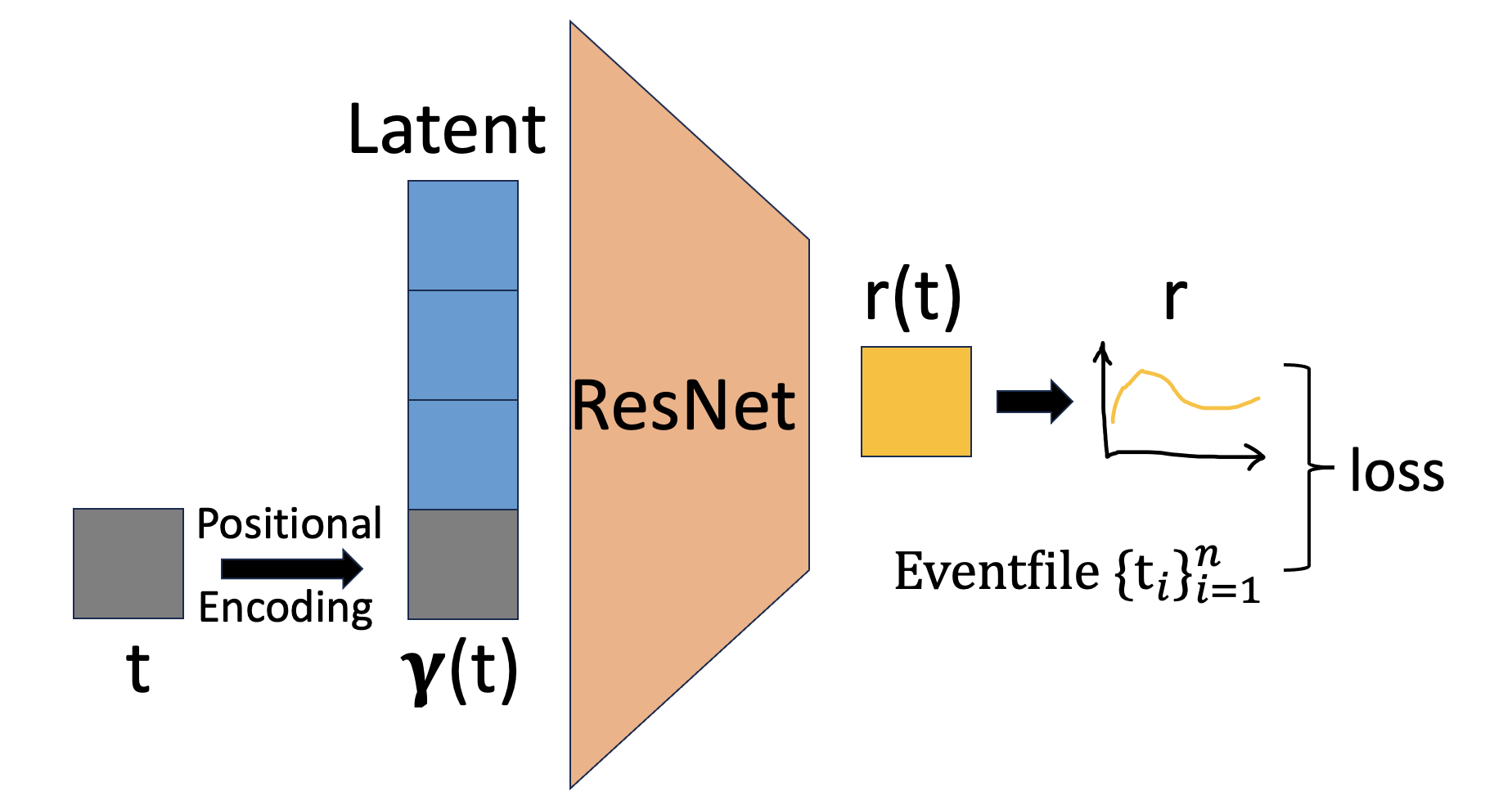}
\caption{Illustration of PPAD. Latent vectors are concatenated to positionally encoded time $t$ and fed to the shared ResNet together. The network outputs the value $r(t)$ of the rate function at time $t$, which, together with values at other times, yield the reconstructed rate function $r$. The rate function $r$ is then used to compute the loss function in \ref{eqn:loss_final} against the event files. When trained with multiple event files, all event files share the same ResNet weights but each has a different corresponding latent vector. Gradients are back-propagated to both the ResNet and the latents.}
\label{fig:PPAD}
\end{figure}

A diagram of the whole PPAD pipeline is given in Figure \ref{fig:PPAD}.

\section{Experiments \& Discussion}\label{sec:experiments}
\subsection{Rate Function Reconstruction}\label{subsec:reconstruction}
\begin{figure*}[ht]
\centering
\includegraphics[width=0.97\textwidth]{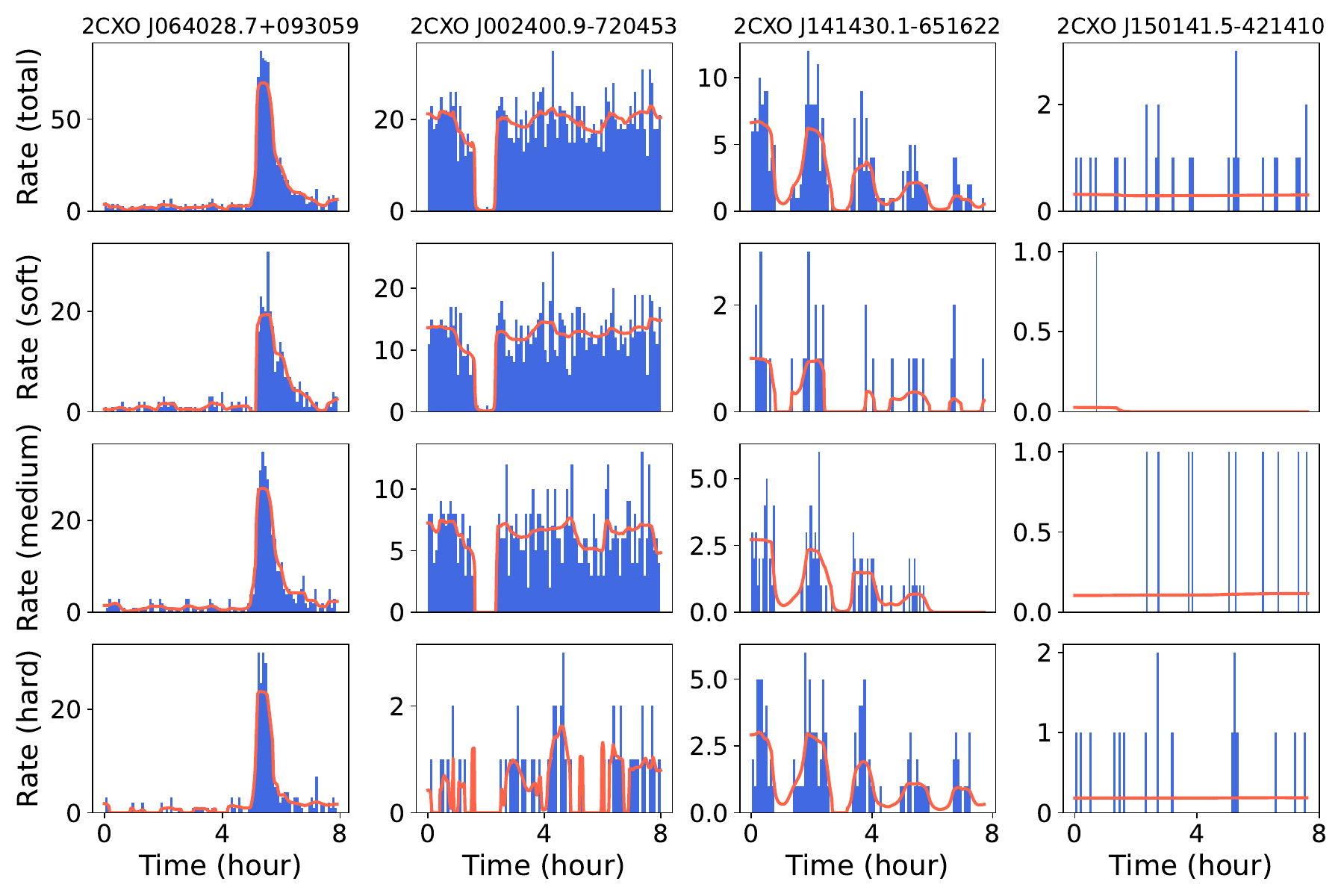}
\caption{Binned event files vs light curves reconstructed by PPAD. Rate from top row to bottom row: total, soft, medium, hard. 
% Names and ObsIDs for respective sources are detailed in Table.~\ref{tab:ids}. 
Event files are binned every $5$ minutes (an arbitrary choice), and reconstructed light curve rates are normalized correspondingly (counts per 5 minutes). Binned event files result in noisy variations. Reconstructed light curves, on the other hand, smooth out the inherit stochasticity of event files while still picking up conspicuous trends.}
\label{fig:reconstruction}
\end{figure*}
PPAD is able to naturally reconstruct X-ray light curves from the event files at any desired resolution. To visualize the quality of light curve reconstruction, Figure~\ref{fig:reconstruction} shows the reconstructed light curves (plotted by sampling on a dense grid of time points) on top of histograms of the raw $28.8$ live kilosecond (ks) event files (binned with a resolution of $0.3$ ks) for a selection of representative sources. We observe that PPAD is able to reconstruct a wide range of light curve shapes, including flares, dips, periodic sources, and sources of constant X-ray flux. The reconstruction quality remains high for the energy-integrated X-ray light curve as well as for specific energy bands, such as the standard soft (0.5~keV-1.2~keV), medium (1.2~keV-2~keV), and hard (2~keV-7~keV) in Chandra observations. The reconstructed light curves are also able to capture transient behaviors, such as the set of astrophysical flares and dips presented in \cite{2024MNRAS.tmp.2687D, dillmann_2025_14589318}, representing phenomena such as type-I X-ray bursts from low-mass X-ray binaries, coronal mass ejections in young stars, and eclipses of ocultation binaries, while smoothing out noisy patterns caused by stochastic photon arrivals. 

Reconstructed light curves for the three energy bands belong to the same event file and therefore share the same latent representation. As a result, information can be shared across energy bands to pick up specific patterns. This is demonstrated, for example, by the soft band of the periodic source shown in Fig.~\ref{fig:reconstruction}. The binned event files resemble those from the low-count source, indicating a possibly constant, non-variable light curve. However, the reconstructed light curve shows periodicity, which is a result of the shared information from other bands where such periodicity is more apparent. Periodicity in certain energy bands can therefore act as a prior that informs the variability in other bands, but the prior is still updated based on the observed photon arrivals.

We note that the exercise we have attempted here does not account for background X-ray photons within the selected aperture of each source. We are not trying to replicate all aspects of light-curve reconstruction, but rather, to understand if a latent representation exists that captures meaningful scientific patterns in X-ray light curves for events of arbitrary duration and number of photon events. However, we will mention that the PPAD method can also be used to recover the background Poisson rate if a background region were selected. Also, in the particular case of Chandra, the low background noise and high spatial resolution imply that for the vast majority of sources, the signal, rather than the background noise, will dominate in the event files.

\subsection{Using the Latent Space: Regression, Classification, and Anomaly Detection}
In addition to light curve reconstruction, PPAD creates a fixed-length vector representation for each event file. In this section, we demonstrate the performance of these learned representations as inputs for downstream tasks, such as source classification and regression on meaningful summary statistics such as spectral hardness and variability. In order to best showcase the rich abstract information contained in these latent vectors, we take a minimalist approach and process them for these tasks using relatively simple machine learning methods. 

\subsubsection{Inferring Source Hardness / Variability}\label{subsec:property}
\begin{figure*}[ht]
    \centering
    \includegraphics[width=\textwidth]{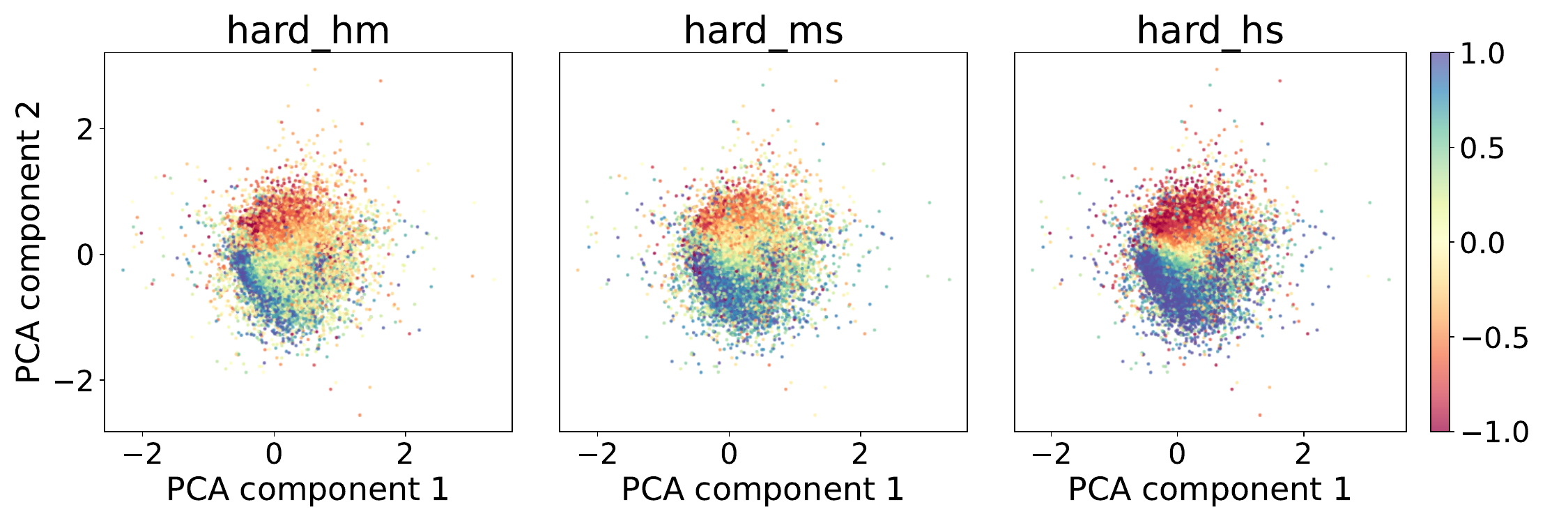}
    \caption{
    Top $2$ principal components of latent features and corresponding hardness ratios. It shows strong relations between the learned representations and meaningful physical features.}
    \label{fig:PCA}
\end{figure*}

Hardness ratios and variability, as summarized in the CSC by properties \verb|hard_hs|, \verb|hard_ms|, \verb|hard_hm|, \verb|var_prob_b|, and \verb|var_index_b|, are important diagnostics of the physical characterization of X-ray sources. For example, hard sources tend to be associated with non-thermal emission related to the acceleration of electrons  in the vicinity of an accreting black hole, such as synchrotron emission; in constrast, soft sources are more likely related to thermal blackbody emission from very hot sources, such as the accretion disk itself. X-ray flux variability, on the other hand, can inform about the timescales of physical processes, such as coronal mass ejections due to magnetic reconnection events in the magnetosphere of young stars, or type 1 bursts in X-ray binaries involving neutron stars.

Therefore, a learned latent representation that codifies hardness and variability is desirable. An important line of previous work in unsupervised X-ray learning uses those properties directly as computed from the CSC for unsupervised and supervised classification. Here we explore if self-supervised learning from the event files themselves can provide an alternative representation that codifies these properties simultaneously. To illustrate that our learned features contain useful information, we explore their relation with the CSC properties.

% First, we visualize the geometry of our learned latent space in Figure~\ref{fig:PCA}, using PCA and tSNE to reduce our dimensionality for ease of exploration. We color-code this representation by computed hardness ratio, variability probability, and variability index. We observe clear continuous trend of hardness ratios. as well as clear clustering of variable sources (medium variability probability, bolometric variability probability, and bolometric variability index) in tSNE space. \Steven{comparison to previous approach would be great again}

In \autoref{fig:PCA}, we visualize the geometry of our learned latent space, using PCA for dimensionality reduction. We color-code this representation by the hardness ratio, as computed from the event files following the prescription of the CSC, and observe a clear continuous trend that hints to the ability of the PPAD to not only reconstruct the light curve, but also to codify the overall spectral shape of the X-ray sources. To confirm this, we use the learned latents to predict the hardness ratio and variability of each source. We do a $80\%-20\%$ train-test split of the data, and then use simple Random Forests with $100$ trees each, which we can use to perform both regression and classification. We use the default hyperparameters in \textit{sklearn} without tuning and performed no cross-validation. For classification tasks, the SMOTE approach \citep{chawla2002smote} with default parameters was applied on the training data to address class imbalance. We summarize results in Table~\ref{tab:supervised_numbers}. In short, we obtain $\sim 0.9$ R2 values on hardness ratio prediction, and $92\%$ accuracy on predicting whether a source is variable (i.e. if its variability index is greater than $5$, indicating variability at a confidence level larger than 90\%). These representations, learned directly from the event files using the PPAD, are valid features for physical characterization of the source, and can be readily obtained for newly observed X-ray sources.

\begin{table}[h]
    \centering
    \begin{tabular}{c c c}
    \hline
        \textbf{Regression Target} & \textbf{MSE} & \textbf{R$^2$}\\
    \hline
        hard\_ms & 0.02 & 0.87 \\
        hard\_hm & 0.01 & 0.88\\
        hard\_hs & 0.01 & 0.94\\
    \hline
        \textbf{Classification Target} & \textbf{Accuracy} & \textbf{F1 Score}\\
        var\_index\_b $> 5$? & 0.92 & 0.63\\
        source type & 0.60 & 0.24\\
        YSO vs AGN & 0.75 & 0.69\\
    \hline
    \end{tabular}
    \caption{Quantitative regression/classification performance of simple models on latent features. All models use $100$ trees with default hyperparameters, are trained on $80\%$ of the data and tested on the remaining set, without cross validation. All numbers are recorded on the test split. The fact that simple predictive model achieve comparable performance as state-of-the-art results (details in Section~\ref{subsec:classification}) demonstrate that latent features are informative representations.}
    \label{tab:supervised_numbers}
\end{table}

\subsubsection{Classifying source types}\label{subsec:classification}
In order to investigate if learned the PPAD latent features also codify information on the astrophysical type of the source, we feed them to a supervised classifier and compare its performance with state-of-the art automatic classification methods. We cross-match our dataset with the labeled set from \citet{yang2022classifying}, which has been curated to provide reliable classes for a large number of CSC sources. This resulted in $5818$ matching X-ray detections\footnote{Note that two or more detections, and therefore two or more even files, might correspond to the same astrophyisical source; this is because we have split long event files into multiple examples, and also because the same source might have been targeted by Chandra more than once.} . We train the classifier in two tasks: \emph{i)} an $8$-label classification between the following types: YSO, AGN, CV, HM-STAR, HMXB, LM-STAR, LMXB, NS, and \emph{ii)} a binary classification between Young Stellar Objects (YSOs) and Active Galactic Nuclei (AGNs). We again make a $80\%-20\%$ train-test split of the data, perform SMOTE to resolve class-imbalance, and use a Random Forest Classifiers with $100$ trees each. As shown in Table~\ref{tab:supervised_numbers}, the $8$-label classification task yields a test accuracy of $60\%$ and a F1 score of $0.24$, and the simpler binary classification (YSO vs AGN) yields a $75\%$ accuracy and a F1 score of $0.69$. 

This comparares fairly with classification approaches that use the CSC properties directly as inputs. For example, \cite{perez2024unsupervised} use clustering-based classification on features selected from prescription approaches, and obtain an average of $61\%$ accuracy on a $4$-label classification task. \cite{yang2022classifying} use a much richer set of features that augment the CSC properties with additional multi-wavelength features such as optical and infrared colors, and perform supervised classification, yielding an $89\%$ accuracy and $0.68$ F1 score on the $8$-label classification task. While a direct comparison is unfeasible due to different data pre-processing methods and models used, the fact that the PPAD embeddings provide accuracies comparable to methods that use pre-computed CSC properties and even multi-wavelength features, imply that the PPAD latents serve as powerful summaries of the astrophyscial properties, and that automatic classification and regression is possible directly from the event files delivered by the observatory.

% I did a test and found optimizing a new latent is actually slower than doing GL algo. Thus we won't make this claim.
% Inferring the latent vectors are fast with a trained PPAD, and such aforementioned predictive power indicates a possible cheaper alternative than features computed from complicated prescription approaches.

% \begin{figure}[ht]
%     \centering
%     \includegraphics[width=0.475\textwidth]{img/tSNE_class.pdf}
%     \caption{tSNE components of latent features with corresponding known class labels from \cite{yang2022classifying}.}
%     \label{fig:tsne}
% \end{figure}

\subsubsection{Anomaly Detection}\label{subsec:anomaly}
\begin{figure*}[ht]
    \centering
    \includegraphics[width=0.97\textwidth]{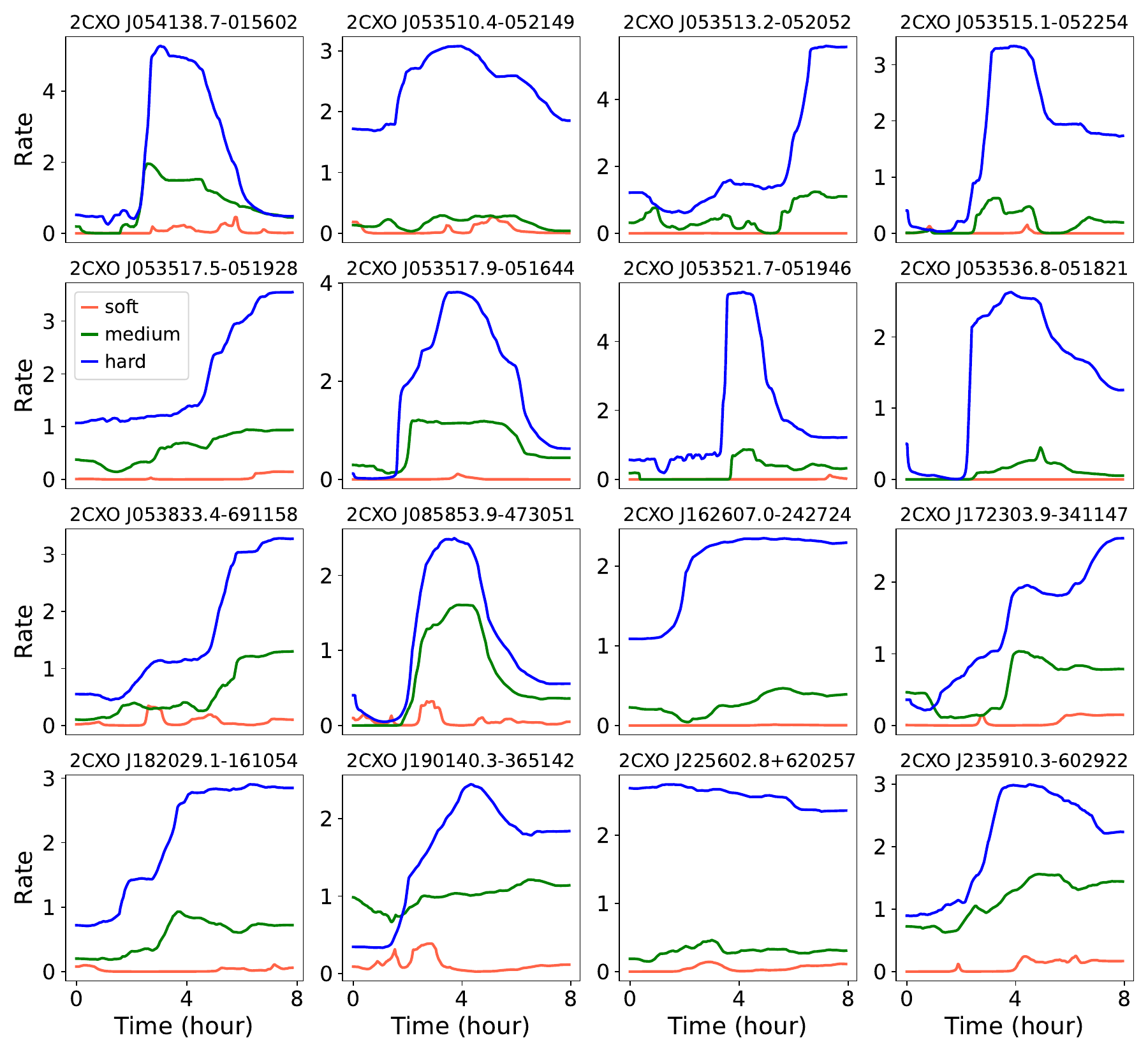}
    \caption{Targeted anomaly (upper left) and $15$ neighboring sources which are closest in the latent space. Almost all found sources are low-count hard-band flares, as the targeted anomaly source does. 
    % Names and ObsIDs for respective sources are detailed in Table.~\ref{tab:ids}.
    }
    \label{fig:anomaly}
\end{figure*}

We perform a simple anomaly search using the learned latents. Among the most interesting detections in the CSC are time-domain anomalies, such as flares and dips in the light curves with particular spectral signatures. For example, a number of relatively soft, fast X-ray transients (FXTs) have been identified in archival searches, that could hint to neutron star mergers or other explosive phenomena \citep{quirola2022extragalactic}. Other flares can be harder, such as those related to magnetic reconnection events in the photosphere of young stars. These can be faint, resulting in low count event files.

To investigate the suitability of the PPAD latents for the identification of anomalies, we select a dim, hard flaring source (2CXO J054138.7-015602) and search for the nearest neighbors of this target in the embedding latent space. Figure~\ref{fig:anomaly} shows PPAD-reconstructed light curves of the target source (upper left) and the $15$ closest neighbors, in the three different three energy bands. We observe that almost all neighboring sources feature low-count, hard-band flares. 

We investigated this further by selecting astrophysical anomalies from the literature and examining their nearest neighbors in the PPAD embedding space. Among the anomalies investigated are eclipsing X-ray binary V*~UY~Vol, a set of FXTs from \cite{lin2022discovery}, and Ultra-Luminous X-ray sources (ULXs). In general, we find that the PPAD embeddings are best at encoding the spectral hardness of the sources (i.e., the neighbors of hard sources are also hard sources), the variability in timescales comparable to the full duration of the observation (i.e., the neighbors of slowly varying light curves are also slowly varying light curves), and the signal-to-noise (i.e., the neighbors of low count detections are also low count detections). Transient phenomena such as flares and eclipsing dips can also be successfully encoded. This demonstrates the potential of PPAD in discovering analogs to interesting time-domain and spectral anomalies, as illustrated by \cite{2024MNRAS.tmp.2687D}, who successfully discover anomalous FXTs using a different representation learning approach.

\subsection{Model Limitations}
Finally, we note some current caveats and limitation of the PPAD model. The first relates to the autodecoder architecture and how it operates at training and test times. Since one needs to prepare a latent vector for every event file, each latent only receives effective gradient updates once per epoch, making autodecoders less efficient than autoencoders during training. More importantly, new latents for unseen data need to be optimized during test time. Although the optimization only takes several seconds, it is still order-of-magnitudes slower than the amortized inference from autoencoders. Introducing an autoencoder that is capable of dealing with variable-length and highly stochastic Poisson arrival times data is a challenging and promising future direction. Relatedly, our current autodecoder architecture is deterministic. An extension to a variational autodecoder may grant a finer control over the distribution of latents.

Another limitation, common in many unsupervised learning pipelines, is the natural trade-off between reconstruction quality and representation quality. In PPAD, this trade-off is controlled by the latent space dimension, the decoder size, and a regularization term. A larger model dictates more focus on reconstruction details, which results in a higher light curve reconstruction quality but less meaningful representations; a smaller model forces learning more abstract and high-level features, therefore resulting in better representations but worse light curve reconstruction. In our experiments, we only ablated the latent dimension. We set the dimension to $8$ after observing that a dimension of $4$ has obviously worse reconstruction quality and a dimension of $16$ leads to worse downstream task performances. A broader exploration of hyperparameters (both in our autodecoder and in the simple random forests used for downstream tasks) can likely strike the balance between these paradigms. Another special parameter that we roughly tuned is the smoothness penalty, and an ideal penalty level should strike a good balance between learning physically meaningful variations and filtering out stochasticity of photon arrivals. %A more careful investigation of specialized network architectures, hyperparameters and training schemes will help researchers targeting more specific tasks. Furthermore, we used simple random forests for our prediction tasks. Better models and more sophisticated tuning will likely result in better downstream task performances.

Finally, event files in our training data are recorded at different starting times and hereafter truncated to $8$ hour segments. This results in variations in the phase of reconstructed light curves and therefore variations in the learned latents. For example, early, mid and late flares have different learned representations, but this difference is likely an artifact of event file recording/truncation and they may in fact come from very similar sources. Designing a phase-shift invariant extension of PPAD to resolve this problem is an exciting future direction. Similarly, to put an even greater focus on variability behaviors like transients, one could design total-count and lifetime invariant extension of PPAD that normalizes event files based on total-counts and lifetimes. As an example, \cite{2024MNRAS.tmp.2687D} normalizes the lifetimes of all event files before computing histograms, which likely encourages the model to focus on variability behaviors, and results in clustering of transient sources in the latent space. Incorporating similar invariance in PPAD would greatly increase the flexibility of the framework by bypassing the truncation and include event files of different lifetimes.

\section{Conclusion}\label{sec:discussion}
A learned representation of X-ray sources that: \emph{i)} results in physically meaningful embeddings; \emph{ii)} can take as input an event file of varying length; and \emph{iii)} accounts for the Poisson nature of the photon-counting process, has been elusive, preventing us from designing effective methods of automatic classification and anomaly detection. Here, we have presented a Poisson Process AutoDecoder (PPAD), a novel framework for end-to-end unsupervised method to encode X-ray sources from their event files. PPAD makes the following key contributions:
\begin{itemize}
    \item It combines the Poisson likelihood function with a total variation penalty as the loss function, thereby yielding light curve reconstructions that not only respect the stochastic nature of Poisson photon arrivals, but also satisfy smoothness constraints.
    \item It proposes to parametrize light curves as one-dimensional neural fields, and applies the Positional Encoding (PE) technique to increase the effective capability of method to capture complex behavior. Additionally, this ensures unlimited resolution of reconstructed light curves as well as natural compatibility with gradient descent algorithms.
    \item Besides reconstructing light curves, it also learns fixed-length latent vectors as abstract representations of event files. These latent representations contain rich information about corresponding X-ray sources and are useful for various downstream tasks.
\end{itemize}

Combining these points, PPAD simultaneously reconstructs light curves and learns latent representations in an end-to-end and unsupervised manner. We verify the efficacy of PPAD in a series of proof-of-concept experiments including light curve reconstruction, source property prediction, source type classification and anomaly detection. PPAD offers a novel way to analyze large quantities of X-ray data (and, more broadly, time series data in the Poisson limit). 

\section*{Acknowledgments}
We extend our gratitude to Vinay Kashyap, Xiao-li Meng, Samuel Perez-Diaz and Carol Cuesta Lazaro for insightful discussion and comments. This research has made use of data obtained from the Chandra Source Catalog provided by the Chandra X-ray Center (CXC). The Villar Astro Time Lab acknowledges support through the David and Lucile Packard Foundation, National Science Foundation under AST-2433718, AST-2407922 and AST-2406110, as well as an Aramont Fellowship for Emerging Science Research. This work is additionally supported by NSF under Cooperative Agreement PHY-2019786 (The NSF AI Institute for Artificial Intelligence and Fundamental Interactions, \url{http://iaifi.org/}).

\bibliography{refs}{}
\bibliographystyle{aasjournal}

\appendix

\section{Implementation Details}\label{sec:implementation}

In this section we provide all implementation details of PPAD, including the neural network architecture, the hyperparameters, the training procedure and details on downstream task experiments.

\subsection{Network architecture} The ResNet takes a $(d_{\text{latent}}+2L+1)$-dimensional input with $d_{\text{latent}}=8$ and $L=12$. It maps the input to a $512$-dimensional hidden vector via a fully connected input layer. The hidden vector is then passed through $5$ fully connected ResNet blocks, maintaining dimensionality. Lastly, a fully connected output layer maps the hidden vector to the output of dimension $3$, representing light curve value at $K=3$ energy bins. 

Each ResNet block has the form $\bm{\Phi}(\bm{x}) = \bm{W}_2 \cdot \sigma(\bm{W}_1 \cdot \sigma(\bm{x})) + \bm{W}_{\text{skip}}\cdot \bm{x}$, where $\bm{W}_1, \bm{W}_2, \bm{W}_{\text{skip}} \in \mathbb{R}^{512\times 512}$ are fully connected layers, and $\sigma$ is the ReLU activation function.

\subsection{Loss function} For hyperparameters in Eqn.~\ref{eqn:loss_final}, we used $\lambda_{\text{TV}}=10,\lambda_{\text{latent}}=1$. The time interval $[0,T)$ with $T=8$ hours is divided into $2048$ evenly spaced bins when we calculate the integral from $\mathcal{L}_{\text{neg-loglikelihood}}$ and a part of $\mathcal{L}_{\text{TV}}$.

\subsection{Training} The training is divided into the following $3$ stages. 

For stage $1$, we create a smaller dataset with higher signal-to-noise ratios. This is done by filtering out many low-count and possibly homogeneous event files, which is the majority of all event files. We remove an event file with probability $1 / (1+\exp(900^{0.99} \cdot n^{0.01} - 900))$, where $n$ is the length (number of photon arrivals) of the event file. The filtering effectively removes mostly low-count event files, and resulted in a higher quality dataset of size $14891$. We then train both the network and corresponding $14891$ latents using the filtered high quality dataset for $1200$ epochs. 

For stage $2$, we switch to the full dataset of size $109656$, but freeze the network and only train the newly added latents for $200$ epochs, in order to provide a good initialization. 

For stage $3$, we again train both the latents and the network together for $600$ epochs. 

We use the Adam optimizer \citep{kingma2014adam} with default hyperparameters for all stages. The learning rate for the latents is 1e-3 for Stages $1 \& 2$ and 1e-4 for Stage $3$. The learning rate for network weights is always $1/10$ of that for the latents. We use a batch size of $64$. The whole training takes approximately $5$ days on a single Nvidia V100 GPU.
\end{document}